%% file: augment_main.tex
\documentclass{article}

\usepackage{PRIMEarxiv}

\usepackage{graphicx,amsmath,url,amsfonts,bm,subcaption,booktabs,nccmath}
\usepackage[mathscr]{eucal}
\usepackage{hyperref,cleveref}
\hypersetup{colorlinks=true,linkcolor=[rgb]{0.1,0.3,0.9},urlcolor=[rgb]{0.1,0.3,0.9},citecolor=[rgb]{0.1,0.3,0.9}}

\title{Data Augmentation-Based Unsupervised Domain Adaptation In Medical Imaging}

\author{
Sebastian Nørgaard Llambias, Mads Nielsen, Mostafa Mehdipour Ghazi \\
Pioneer Centre for AI, Department of Computer Science \\
University of Copenhagen \\
Copenhagen, Denmark \\
\texttt{\{snl,madsn,ghazi\}@di.ku.dk} \\
}

\begin{document}
\maketitle

\begin{abstract}
\input{texts/0_abstract}
\end{abstract}


\section{Introduction}\label{sec:intro}
\input{texts/1_introduction}

\section{Method}\label{sec:method}
\input{texts/2_method}

\section{Experiments and Results}\label{sec:experiment}
\input{texts/3_experiments_results}

\section{Conclusion}\label{sec:conclusion}
\input{texts/4_conclusion}

\section*{Acknowledgments}
This project has received funding from Innovation Fund Denmark under grant number 1063-00014B, Lundbeck Foundation with reference number R400-2022-617, and Pioneer Centre for AI, Danish National Research Foundation, grant number P1.

\bibliographystyle{unsrt}  
\bibliography{references}

\end{document}

%% file: texts/0_abstract.tex
Deep learning-based models in medical imaging often struggle to generalize effectively to new scans due to data heterogeneity arising from differences in hardware, acquisition parameters, population, and artifacts. This limitation presents a significant challenge in adopting machine learning models for clinical practice. We propose an unsupervised method for robust domain adaptation in brain MRI segmentation by leveraging MRI-specific augmentation techniques. To evaluate the effectiveness of our method, we conduct extensive experiments across diverse datasets, modalities, and segmentation tasks, comparing against the state-of-the-art methods. The results show that our proposed approach achieves high accuracy, exhibits broad applicability, and showcases remarkable robustness against domain shift in various tasks, surpassing the state-of-the-art performance in the majority of cases.

%% file: texts/1_introduction.tex
Domain shift occurs when models are exposed to large distributional shifts between training and testing data. This frequently occurs when models trained on clean academic grade data are applied to workflow-optimized clinical data, and generally results in considerably degraded generalization. To alleviate this effect, models must be adapted to the target domain. In principle, this can be achieved using annotated samples from the target domain, but, in practice, this is rarely feasible due to the cost and scarcity of annotated samples in medical imaging. Therefore, domain adaptation is usually achieved using unsupervised domain adaptation (UDA) methods.

Recent efforts in UDA \cite{Kamnitsas,He,ZUO2021118569} often leverage generative adversarial networks (GANs) to mitigate the lack of annotated target data. Moreover, adversarial approaches leveraging data augmentation have been shown to outperform state-of-the-art adversarial methods not employing data augmentation \cite{Orbes-Arteaga_DA_UDA}.  However, adversarial training is volatile and sensitive to design choices, often relying on dataset-specific hyperparameter tuning and access to samples from the target distribution \cite{Orbes-Arteaga_KD,Kushibar}. These traits render adversarial training undesirable in designing a domain-agnostic method robust to distributional shifts in geometry, demography, imaging sequence, and task.

Available augmentation policies include the generic RandAugment \cite{RandAugment} and the specialized Med-Aug methods \cite{Lo}. RandAugment optimizes all operations jointly with a common magnitude parameter, massively reducing the search space while demonstrating state-of-the-art accuracy. However, the method depends on defining complementary magnitudes, lacks ordering, and allows reoccurring augmentations, which can render medical images unrealistic. Med-Aug \cite{Lo} uses a selection of generic and medical imaging-specific augmentations. It retrieves augmentation policies using an efficient and high-dimensional covariance optimizer but is unable to demonstrate competitive accuracy.

Recent studies in medical image segmentation have often been based on the U-Net architecture \cite{Ronneberger} and its improved variants such as nnU-Net \cite{Isensee} and MultiResUNet \cite{Ibtehaz_2020}. These networks demonstrate state-of-the-art accuracy in a range of tasks manifesting varying degrees of domain shift, but solutions are often specialized to the properties of the datasets, limiting their applicability for endeavors in UDA where the distance between the properties of the source and target datasets can be substantial.

We propose a robust unsupervised domain adaptation pipeline for medical image segmentation exclusively using MRI-specific data augmentation techniques implemented for 2D and 3D and applied online. The novelty of our approach does not lie in making models more robust through augmentation nor in the uniqueness of our augmentations. Rather, the novelty of the proposed pipeline is its demonstrable efficacy and competitive performance in an array of challenging and distinct tasks. The core of the study is a hippocampus segmentation task, serving to demonstrate the superior robustness of our method against the state-of-the-art unspecialized solution. To this end, we use two architectures and four datasets incurring intra-task domain shifts in location, demography, imaging sequence, and network architecture.

Additionally, to validate the robustness of our method against state-of-the-art, we apply our unspecialized solution to the MICCAI grand challenge dataset on multi-domain cross-time-point infant cerebellum MRI segmentation 2022 (cSeg-2022) \cite{cseg22} and the multi-source white matter hyperintensity (WMH) segmentation challenge from MICCAI 2017 \cite{Kuijf}, incurring severe domain shifts in demography (from adult brains to infant's brains) and task (from anatomical regions to pathologies). In both cSeg-2022 and WMH challenges our method is competitive with the challenge winners' specialized solutions without using additional sources or processing for training.

%% file: texts/2_method.tex
The proposed UDA pipeline is integrated into the robust and simple nnU-Net \cite{Isensee} framework, both to limit specialization towards specific domain properties and to use our method in a state-of-the-art environment. The pipeline is agnostic to network architecture and trains fully convolutional 2D or 3D networks with extensive online augmentation. To validate that findings are not model-specific, we also employ both MultiResUNets \cite{Ibtehaz_2020} and traditional U-Nets. 

For the hippocampal segmentation task, the basic nnU-Net pipeline serves as the state-of-the-art unspecialized baseline. This pipeline employs exclusively generic data augmentation transforms to simulate common imaging artifacts and variations. This includes scaling, rotation, Gaussian noise and blur, deformation, brightness, contrast clipping, low-resolution simulation, gamma correction, and mirroring. Likewise, the proposed pipeline utilizes MRI-specific transforms presented in \cite{fast-aid} for vastly increased robustness in MRI segmentation tasks alongside the generic transforms.

\subsection{U-Net Architectures}

The U-Net-like architecture largely resembles the original U-Net, only diverging by training with deep supervision, which computes additional losses from outputs of deeper layers in the decoder using a downsampled version of the ground truth \cite{Isensee}. It is implemented with instance normalization and the stochastic gradient descent (SGD) optimizer. The MultiResUNet is distinguished by its residual path, replacing the traditional skip-connections with a sequence of convolutions and residual connections \cite{Ibtehaz_2020}. It is implemented with batch normalization and the Adam optimizer.

\subsection{Data Augmentation}\label{subsec:dataaug}

\begin{figure}[ht]
\centering
\includegraphics[width=\linewidth]{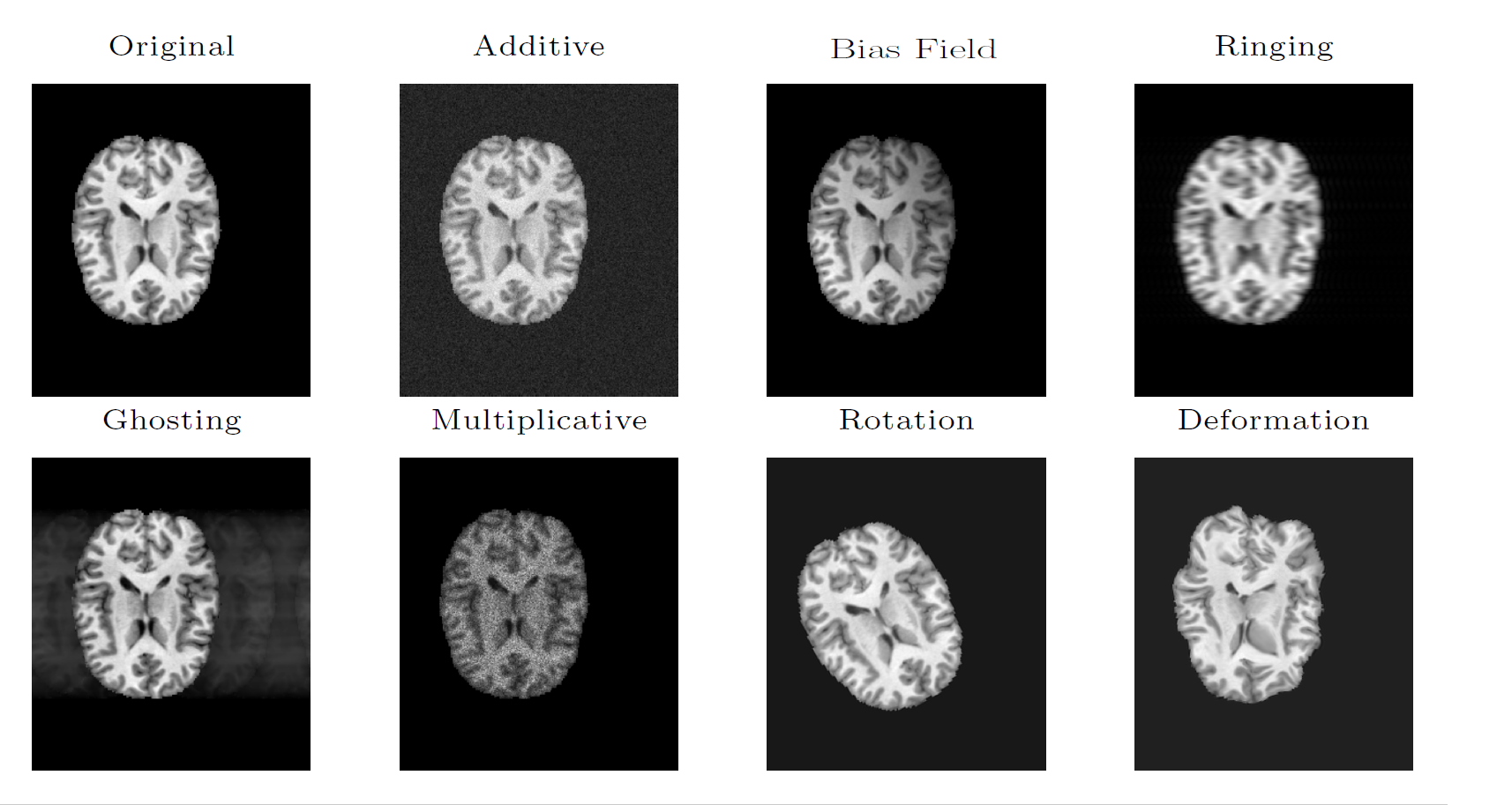}\caption{Exaggerated augmentation samples obtained in 3D and visualized in 2D on the same axial slice}
\label{fig:artefacts}
\end{figure}

The augmentation techniques used in the proposed pipeline are additive noise, multiplicative noise, intensity inhomogeneity distortion (bias field), rotation, elastic deformation, and Gibbs ringing and motion ghosting artifacts. Augmentations are obtained in 3D to generate realistically transformed MRI volumes. Figure~\ref{fig:artefacts} shows examples of the exaggerated augmentations visualized on a single axial slice.

\textit{Noise:} Gaussian (additive) and speckle (multiplicative) noise is common to MR images. We emulate the additive noise with $\mu = 0$ and $\sigma \in$ [0, 0.0001] and the multiplicative noise with $\mu = 0$ and $\sigma \in$ [0, 0.001].

\textit{Intensity Inhomogeneity Distortion:} Intensity inhomogeneity distortion or bias field is a shading effect that appears over the image. Bias fields are applied with random 3D centers.

\textit{Rotation:} Rotation is performed as the head orientation can differ even in the same position. Volumes are rotated along all three axes by a degree in [-30, 30], after positioning the head in the RAS orientation.

\textit{Elastic Deformation:} Randomly deformed shapes of volumes are generated by interpolating random 3D displacement fields smoothed by a Gaussian kernel with $\sigma \in$  [20, 30] and scaled by a multiplicative factor $\alpha \in$ [200, 600].

\textit{Ringing Artifact:} Gibbs ringing artifacts appear as oscillating bands near abrupt signal changes. Implementation is done by applying the fast Fourier transform to the volumes in all three directions and cutting the edges of the k-space at an integer $\in$ [96, 128] along a randomly chosen axis.

\textit{Ghosting Artifact:} Motion ghosting artifacts appear as repeated ghosts of brains. The artifact is implemented by weighing every \textit{n}-th point in the k-space by a factor $\in$ [0.85, 0.95] where the integer \textit{n} $\in$ [2, 10] represents the number of ghost brains.

%% file: texts/3_experiments_results.tex
\subsection{Study Data}

The hippocampal segmentation task includes 240 T1-weighted MRI scans from four independently collected and annotated datasets \textit{Hammers} \cite{Faillenot,Gousias,Hammers}, \textit{HarP} \cite{Boccardi}, \textit{LPBA40} \cite{Shattuck}, and \textit{OASIS} \cite{Landman}. Additional details on the datasets are shown in Table \ref{tab: data}. The datasets cover a range of acquisition parameters, labeling techniques, and demographics. Hammers, LPBA40, and OASIS include only cognitively normal subjects while HarP includes cognitively normal and impaired subjects and patients with Alzheimer’s disease. The high degree of intra-task data heterogeneity serves to avoid overfitting findings to specific scanner vendors, scanner strengths, age groups, and nationalities.
\input{illustrations/dataset_info}

Subsequently, to demonstrate the applicability of the proposed pipeline on non-T1 sequences and non-hippocampal segmentation tasks, we include the \textit{WMH} \cite{Kuijf} and the \textit{cSeg-2022} \cite{cseg22} challenge datasets. The WMH dataset is a multi-site and multi-vendor white matter lesion segmentation task consisting of 2D multi-slice FLAIR scans. The cSeg-2022 consists of a publicly available training set ($N=13$) of 24-month-old infants and an unreleased test set ($N=20$) of 24-month and 6-month-old infants. Both partitions of the cSeg-2022 dataset contain T1-weighted brain MRI scans but only the test set includes 6-month-old infants. This makes the generalization task very challenging due to the considerable volumetric and shape change associated with the age disparity.

\subsection{Setup}

In UDA we are generally oblivious to the biases distancing the target distribution from the source. Therefore, to increase the likelihood of modeling relevant biases, we introduce as much variance as possible using our augmentations without deteriorating models. To determine the maximum non-deteriorating augmentation parameters we conducted experiments with shorter training schemes to assess both the maximum augmentation frequency and magnitude. To estimate the maximum augmentation frequency we trained 6 models with the augmentation probabilities $p_{aug} = [0, 0.1, 0.2, 0.3, 0.4, 0.5]$. That is, for $p_{aug} = 0.2$ there is a 20\% probability to apply each of the augmentations, often subjecting samples to multiple augmentations. We observed significantly ($p < 0.001$ using paired t-test) higher Dice similarity coefficient (DSC) for models with $p_{aug} \in [0.3, 0.4]$. Based on this we set $p_{aug} = 1/3$ as the default frequency.

To estimate the optimal magnitude we defined 5 magnitudes for each of the 7 augmentations, centered around the values proposed in \cite{fast-aid}. For each of the 35 magnitudes, we trained a model on OASIS $(N=10)$ and evaluated its intra- and inter-domain generalization on OASIS $(N=5)$ and HarP $(N=100)$. We found significant positive Pearson correlations between magnitude and DSC on the HarP samples for bias field ($r = 0.73, p < 0.001)$, motion ghosting ($r=0.72, p < 0.001$), and rotation ($r=0.90, p < 0.001)$, while only observing very limited degradation on the OASIS validation. Gibbs ringing resulted in minor degradation, plausibly because the datasets are so clean it renders volumes unrealistic. Correlations were not found for the remaining augmentations. However, that is to be expected as the primary purpose of our noise injections and deformation is to facilitate training for extended periods without overfitting. The values reported in Section \ref{subsec:dataaug} were based on these experiments.

Furthermore, an ablation study was conducted by training models with all but one augmentation. The study revealed noise injections and deformation to be of limited impact and the remaining augmentations to be of larger impact with bias field and rotation transforms being the most important. Thus, further supporting the notion that noise injections and deformation inflate the dataset with tolerable deterioration, while the remaining augmentations bridge the gap between the distributions.

\subsection{Results}

With the augmentation parameters obtained from the mentioned experiments, we trained a series of models for the hippocampus, white matter lesion, and cerebellum segmentation tasks using the U-Net and MultiResUnet network architectures in both 2D and 3D. In the hippocampus task, we used the state-of-the-art nnU-Net \cite{Isensee} framework as the baseline. This serves to assess the proposed unspecialized UDA pipeline against the state-of-the-art unspecialized solution in medical image segmentation. We employed both U-Nets and MultiResUnets models to certify findings were not architecture-specific. First, to compare robustness in a data-rich setting, we trained multi-source MultiResUnets on the Hammers, HarP, and LPBA40 training sets ($N=150$) and applied them to the four hippocampal test sets. Afterward, to compare robustness in a data-scarce setting, we trained single-source U-Nets on the OASIS training set ($N=15$) and applied them to the four hippocampal test sets.

\input{illustrations/Hippo}

The results of the hippocampus task are presented in Table~\ref{tab:hippo}. As can be seen, in the data-scarce setting our augmentation policy improves the generalization from OASIS to HarP and LPBA40 by a factor of 2, and in the data-rich setting we see significant improvements for 3/4 test sets. In the less complicated domains, we see a negligible effect, but in the difficult domain adaptation tasks we see massive improvements in domain adaptation from as few as 15 samples, substantiating the overall positive impact also observed in \cite{fast-aid}.  

\begin{figure}[t]
    \centering
    \includegraphics[width=0.85\linewidth]{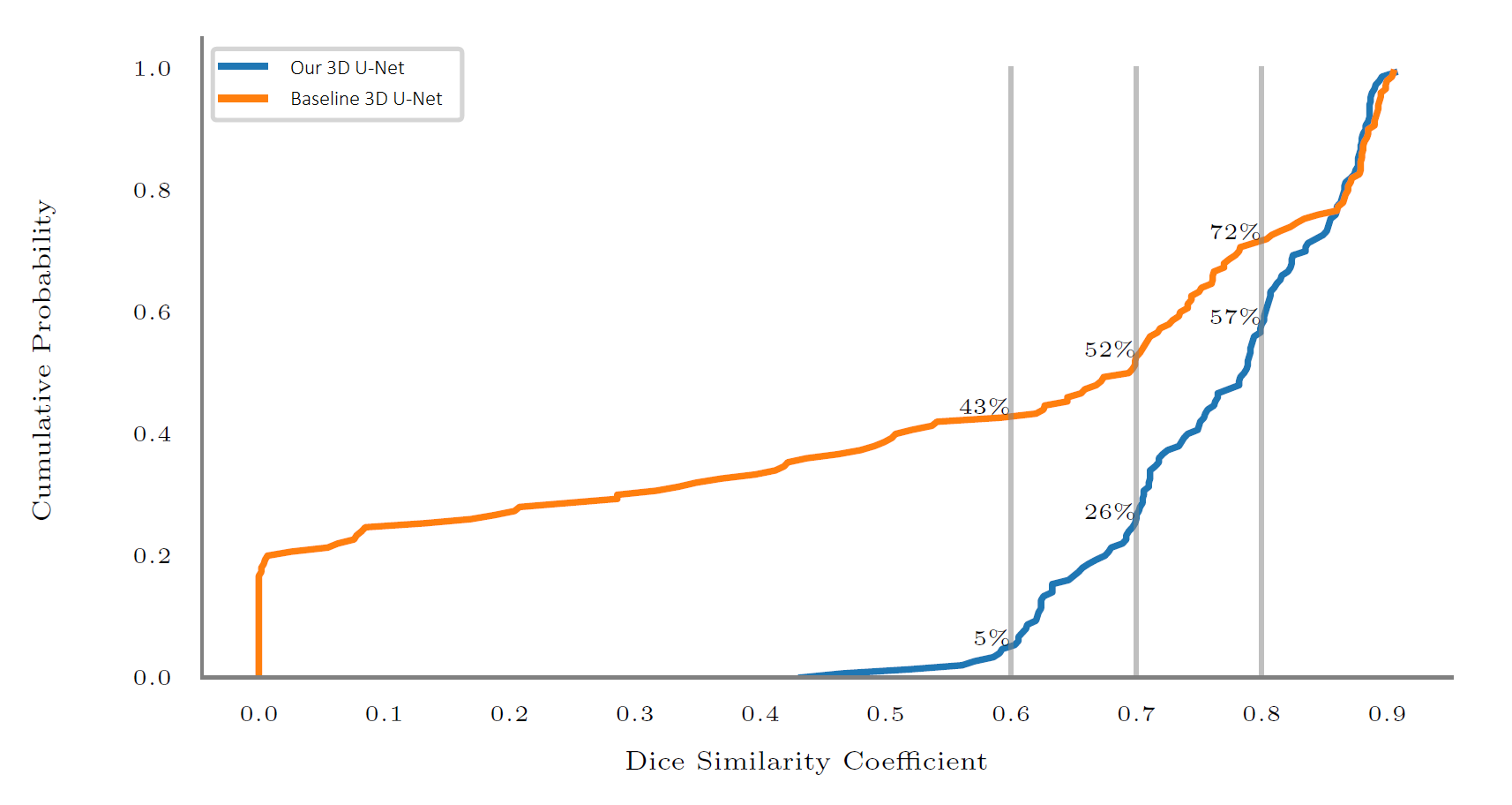}
    \caption{Cumulative distribution of DSC for the baseline and our U-Net models trained on OASIS ($N=15$) and applied to Hammers, HarP, OASIS, and LPBA40.}
    \label{fig:ECDF}
\end{figure}

The improvements observed in Table~\ref{tab:hippo} are not the results of increasing the upper generalization bound. Rather, the proposed pipeline makes the models robust to domain shifts by massively raising the lower generalization bound and eliminating total segmentation failures. This newfound robustness is demonstrated by the drastically reduced standard deviations in Table~\ref{tab:hippo} and illustrated in Fig.~\ref{fig:ECDF}. The figure shows that roughly 20\% of the baseline predictions are total segmentation failures (0.0 DSC) while our model produces none of them. If one assumes a clinical error tolerance of 0.6 DSC, the baseline segmentations will be erroneous 43\% of the time while our method would be wrong in 5\% of the cases. For 0.7 DSC, those values are 52\% to 26\%, and for 0.8 DSC, they are 72\% to 57\%, showing considerable gains in reliability for any tolerance threshold at the cost of virtually no top-end performance.

In the WMH and Cerebellum tasks, we used the challenge winners as the state-of-the-art baseline. This serves to assess the proposed unspecialized UDA pipeline against state-of-the-art specialized solutions. We employed 2D and 3D models to certify findings were not dimensionality-specific. Moreover, to certify findings were not sequence-specific, our model was exclusively trained on the FLAIR images in the WMH task whereas the challenge winner's model was trained on both FLAIR and T1 modalities. The results of these tasks are presented in Table~\ref{tab:wmh_cseg}. In the WMH task, we did not find a significant difference ($p < 0.001$ using a one-sample t-test) between the DSC of our unimodal model and that of the challenge winner's specialized multimodal model. Likewise,  in the cSeg-2022 challenge, we did not find significant differences between our models and those of the challenge winners in the 24-Month and 6-Month tasks. These challenge tasks demonstrate the out-of-the-box competitiveness and robustness of our solution versus state-of-the-art solutions specialized for the tasks at hand using preprocessing tailored to the training data, hyperparameter tuning, large model ensembles, and task-specific post-processing \cite{LI}.

\input{illustrations/WMH_cSeg}

%% file: illustrations/dataset_info.tex
\begin{table}[b]
\caption{Overview of the datasets used in this study. Country codes are ISO Alpha-2 and vendor abbreviations refer to GE (G), Philips (P), and Siemens (S).}
\centering
\resizebox{0.85\linewidth}{!}{%
\begin{tabular}{lcccccc}
\toprule
\textbf{Dataset} & \textbf{Task} & \textbf{Train/Test} & \textbf{Age} & \textbf{Site} & \textbf{Strength} & \textbf{Vendor}   \\
\midrule
Hammers \cite{Faillenot,Gousias,Hammers} & Hippocampus & 20/10 & 20-54 & UK & 1.5T & G \\
HarP \cite{Boccardi} & Hippocampus & 100/35 & 55-90 & CA, US & 1.5-3T & G, P, S \\
LPBA40 \cite{Shattuck} & Hippocampus & 30/10 & 19-40 & US & 1.5T & G \\
OASIS \cite{Landman} & Hippocampus & 15/20 & 18-90 & US & 1.5T & S  \\
WMH \cite{Kuijf} & Brain Lesion & 60/110 & - & NL, SG  & 1.5-3T & G, P, S \\
cSeg-2022 (24M) \cite{cseg22} & Cerebellum & 13/5 & 2 & US & - & - \\
cSeg-2022 (6M) \cite{cseg22} & Cerebellum & 0/15 & 0.5 & US & - & - \\
\bottomrule
\end{tabular}
}
\label{tab: data}
\end{table}

%% file: illustrations/Hippo.tex
\begin{table}[b]
\centering
\caption{DSC (mean$\pm$SD) for test segmentations. Significant differences ($p < 0.001$ using paired t-test) are boldfaced. Model names signify network dimensionality (\textbf{2}D or \textbf{3}D), architecture (\textbf{M}ultiResUNet or \textbf{U}-Net), and source data. For instance, 3U$^{4}$ refers to a 3D U-Net trained on the OASIS$^{4}$ training data.}
\begin{tabular}{lcccc}
\toprule 
 Model & Hammers$^1$ & HarP$^2$ & LPBA40$^3$ & OASIS$^4$ \\
\midrule
Ours 3M$^{123}$ & $0.84 \pm 0.02$ & $\textbf{0.88} \pm 0.07$ & $\textbf{0.84} \pm 0.04$ & $\textbf{0.77} \pm 0.04$ \\
Base 3M$^{123}$ 
& $0.85 \pm 0.03$ & $0.72 \pm 0.28$ & $0.72 \pm 0.21$ & $0.58 \pm 0.20$ \\
\midrule
Ours 3U$^4$ & $0.69 \pm 0.04$ & $\textbf{0.75} \pm 0.08$ & $\textbf{0.63} \pm 0.03$ & $0.88 \pm 0.01$\\
Base 3U$^4$ & $0.70 \pm 0.05$ & $0.37 \pm 0.32$ & $0.27 \pm 0.25$ & $0.83 \pm 0.19$\\
\bottomrule
\end{tabular}
\label{tab:hippo}
\end{table}

%% file: illustrations/WMH_cSeg.tex
\begin{table}[b]
\centering
\caption{Average DSC for test segmentations. No significant differences ($p < 0.001$ using a one-sample t-test) were found. Model names signify network dimensionality (\textbf{2}D or \textbf{3}D), architecture (\textbf{M}ultiResUNet or \textbf{U}-Net), and source data. For instance, 2M$^{1}$ refers to a 2D MultiResUnet trained on the WMH$^{1}$ training data.}
\begin{tabular}{lccc}
\toprule 
Model & WMH$^1$ & cSeg 24-Month$^2$ & cSeg 6-Month$^3$ \\
\midrule
Ours (2M$^1$) & 0.79 & - & - \\
Challenge Winner & 0.80 & - & - \\
\midrule
Ours (3U$^2$) & - & 0.91 & 0.78 \\
Challenge Winner & - & 0.93 & 0.79 \\
\bottomrule
\end{tabular}

\label{tab:wmh_cseg}
\end{table}




%% file: texts/4_conclusion.tex
In this paper, we proposed an extensive 3D augmentation policy for robust unsupervised domain adaptation in medical imaging using MRI-specific data augmentation techniques. Our models benefited from a large degree of variations introduced during the training and successfully generalized to unseen target domains from very few source samples.

Our findings span several segmentations tasks, demographics, scanner vendors and strengths, MRI sequences, network architectures, and model dimensionality to validate the robustness and reliability of the proposed pipeline, both against state-of-the-art unspecialized solutions and even specialized solutions. Additionally, the proposed method is adopted in retrospective observational studies using data from different sites including the capital region of Denmark comprising more than 100,000 brain MRI scans to segment brain structures and determine volumetry changes, as well as to detect white matter lesions \cite{schiavone2023robust} and cerebral microbleeds \cite{ferrer2023deep}.